# Beyond Beats: A Recipe to Song Popularity?
## A machine learning approach

Niklas Sebastian Jung, Florian Mayer

University of Innsbruck

Email: niklas.jung@student.uibk.ac.at, florian.mayer@student.uibk.ac.at

**Abstract**

Music popularity prediction has garnered significant attention in both industry and academia, fuelled by the rise of data-driven algorithms and streaming platforms like Spotify. This study aims to explore the predictive power of various machine learning models in forecasting song popularity using a dataset comprising 30,000 songs spanning different genres from 1957 to 2020. **Methods:** We employ Ordinary Least Squares (OLS), Multivariate Adaptive Regression Splines (MARS), Random Forest, and XGBoost algorithms to analyse song characteristics and their impact on popularity. **Results:** Ordinary Least Squares (OLS) regression analysis reveals genre as the primary influencer of popularity, with notable trends over time. MARS modelling highlights the complex relationship between variables, particularly with features like instrumentalness and duration. Random Forest and XGBoost models underscore the importance of genre, especially EDM, in predicting popularity. Despite variations in performance, Random Forest emerges as the most effective model, improving prediction accuracy by 7.1% compared to average scores. Despite the importance of genre, predicting song popularity remains challenging, as observed variations in music-related features suggest complex interactions between genre and other factors. Consequently, while certain characteristics like loudness and song duration may impact popularity scores, accurately predicting song success remains elusive.

# Introduction

Music holds a profound significance in human existence, serving as a universal language that transcends cultural barriers and speaks directly to our emotions and experiences. From ancient civilizations to modern societies, music has played a central role in rituals, celebrations, storytelling, and personal expression (Mithen, 2009). One of the key reasons music is important to us humans is its ability to evoke emotions and memories. Whether it's the joy of a catchy tune, the melancholy of a soulful melody, or the adrenaline rush of a powerful beat, music has a unique power to affect our mood and transport us to different times and places. Furthermore listening to music helps us humans with mood regulation (Lonsdale & North, 2011).

In recent years, the advent of streaming platforms like Spotify has revolutionized the way we consume and interact with music. These platforms have made music more accessible than ever before, allowing users to discover, explore, and enjoy a vast library of songs from all genres and eras with just a few clicks. This fundamentally changed the landscape of music and especially popular music, as nowadays discovering new music is not like it used to be, by friends recommending specific albums or artists. Music consumer behaviour changes in a direction in which data-driven algorithms play a more important role when it comes to suggesting new and popular music.

These algorithms analyse users' listening habits, preferences, and interactions to tailor personalized recommendations, playlists, and discover new artists. By leveraging data on listening history, likes, shares, and other user behaviours, Spotify's algorithms curate Discover Weekly playlists and Release Radar, offering users a customized selection of tracks each week based on their tastes. This personalized approach not only introduces listeners to new music but also creates a sense of serendipity and excitement as users stumble upon unexpected gems that resonate with them (Bello & Garcia, 2021). However, while algorithms have undoubtedly enhanced the music discovery experience, they also raise concerns about the homogenization of taste and the potential for echo chambers. By prioritizing recommendations based on past preferences, algorithms may inadvertently reinforce existing biases and limit exposure to diverse perspectives and genres, even though evidence has been found, that motives and genres change as people get older. (Lonsdale & North, 2011)

As Spotify has become more data driven, music itself has also become more an object which can be seen as aggregated data points such as how vocal is a song or how popular is a particular artist. This shift towards quantifying musical attributes has not only transformed how music is analysed and categorized but has also influenced the creation process itself. Artists and producers now often consider the data-driven preferences of listeners, shaping their compositions to align with trends and maximize their appeal in the digital realm. (Hodgson, 2021)

But is this really possible? This is not just the question we are asking ourselves; it has been asked multiple times by labels, artists, listeners and last but not least us – data scientists. Can music sufficiently be broken into data bits and do these add up to explain song popularity?

To investigate this question, we utilize simple machine learning algorithms with a large data set, to see if we can find general patterns of song characteristics that can predict if a song will become be popular or not.

## Related Literature

Surprisingly there is already a lot of research on this specific topic. So much even that apparently it exists an independent research field with its own name, "Hit Song Science" (HHS). (Ioannis Dimolitsas et al., 2023) So not just labels and music vendors are obsessed with this topic, data scientists are as well, leading to countless publications. One particular reason for this might be that Spotify itself offers free developer account with a Web API, that supports gathering official song data from Spotify, in a well-prepared manner. (Spotify, 2024) Naturally scientists have tried to analyse aggregated song data (See Agha Haider Raza & Krishnadas Nanath, 2020; E. Georgieva et al., 2018; Joshua S. Gulmatico et al., 2022; Saragih, 2023).

Glumatico and colleagues conclude that linear models don't lead to satisfying results as the relationship between musical characteristics and popularity cannot be described with a linear relationship. Therefore machine learning algorithms that can capture non-linearity are more promising. (Joshua S. Gulmatico et al., 2022)

Different approaches have been tried, Dimolitsas and colleagues divided between hits and non-hits, for hits reaching Top100 billboard in a certain year. They then created a balanced data set with an equal amount of hits and non-hits, to then use Random-Forrest or Support-Vector-Machine. They were able to predict hits with a accuracy between 62-75% (Ioannis Dimolitsas et al., 2023)

In general, is the prediction accuracy, a priori relatively weak. Eva Zangerle achieved better results when inspecting a posteriori property (after song release), like promotions, social media strategy and user perception. Zangerle was successful at predicting future chart hits analysing contemporary Tweets. (Zangerle, 2016)

The machine learning models that Raza and Nanath created did not improve prediction accuracy significantly compared to a coin toss. Leading them to conclude that "there is no magic formula for predicting song success" (Agha Haider Raza & Krishnadas Nanath, 2020)

A statement we're eager to put to the test! However, in all honesty, unravelling such a formula would diminish some of the enchantment and passion inherent in the music experience, so we wouldn't be too upset when, in our research attempt, such a formula would remain concealed.

## Our Data

We utilized the data set created by Joakim Arvidsson from Kaggle.com (Arvidsson, 2023). He retrieved the data through the Spotify Web API and by using the R-package "spotifyr" (Thompson et al., 2022) to create this data set which contains 30.000 songs from Spotify (including duplicates). The songs are from 6 different genres, including EDM, Latin, Pop, R&B, Rap and Rock from 1957 until 2020.

This dataset comprises a diverse array of variables capturing essential attributes of songs and playlists.

Beginning with song-specific details, the dataset includes identifiers such as the unique track ID, name of the song, and the artist(s) responsible for its creation. Additionally, it provides insights into the song's popularity, measured on a scale from 0 to 100, and links each track to its respective album via an album ID and name. Furthermore, the dataset records the release date of the album housing the song.

Moving to playlist characteristics, the dataset offers information about playlist names and their corresponding unique IDs, along with genre and subgenre classifications, aiding in the categorization and organization of songs within playlists. The dataset also delves into the musical attributes of each track. From danceability and energy, which gauge a song's suitability for movement and its intensity, respectively, to more nuanced features such as key, loudness, and mode, which shed light on the song's tonal quality and structure. Further, the dataset includes measures like speechiness, acousticness, and instrumentalness, providing insights into the presence of spoken words, acoustic elements, and vocal content within the songs. Additionally, attributes like liveness and valence offer indications of whether a recording captures live performance energy and the overall emotional tone of the music. Finally, the dataset incorporates tempo and duration, offering quantitative assessments of a song's pace and length, crucial for understanding its rhythm and temporal characteristics.

To help visualizing these variables we introduce two, somewhat contrary, examples of how musical experiences can be described by these variables:

"Listen Before I Go" by Billie Eilish takes listeners on a journey characterized by its slow tempo and extended duration. With subdued loudness and a predominance of acoustic elements, this track delves deep into raw emotion, offering an intimate listening experience. Its valence, reflecting the emotional tone, tends towards the lower end, evoking feelings of sadness and introspection.

On the other hand, "Don't Stop Me Now" by Queen, bursts onto the scene with its rapid tempo and infectious energy. The song's dynamic rhythm and lively instrumentation propel listeners into a whirlwind of excitement and movement. With high energy and danceability, it's a magnetic force on the dance floor, radiating positivity and joy. Furthermore, its valence is elevated, exuding an exuberant and uplifting vibe that's hard to resist.

One notable aspect of our dataset is the distribution of songs across various genres. We observe a relatively similar number of songs within each genre category, indicating a balanced representation of musical styles. Specifically:

> Rap: 3968 songs
> Pop: 3839 songs
> Electronic Dance Music (EDM): 3389 songs
> R&B: 3235 songs
> Rock: 3127 songs
> Latin: 3030 songs

This even distribution across genres ensures a diverse range of musical styles is accounted for in our analysis, allowing for comprehensive insights into the factors influencing song popularity across different genres.

Furthermore, we present a histogram illustrating the frequency distribution of songs by artists, revealing that the majority of artists are represented once or twice in the dataset. Only a select few artists or bands appear more frequently, with Queen having the highest representation with a maximum of 128 songs.

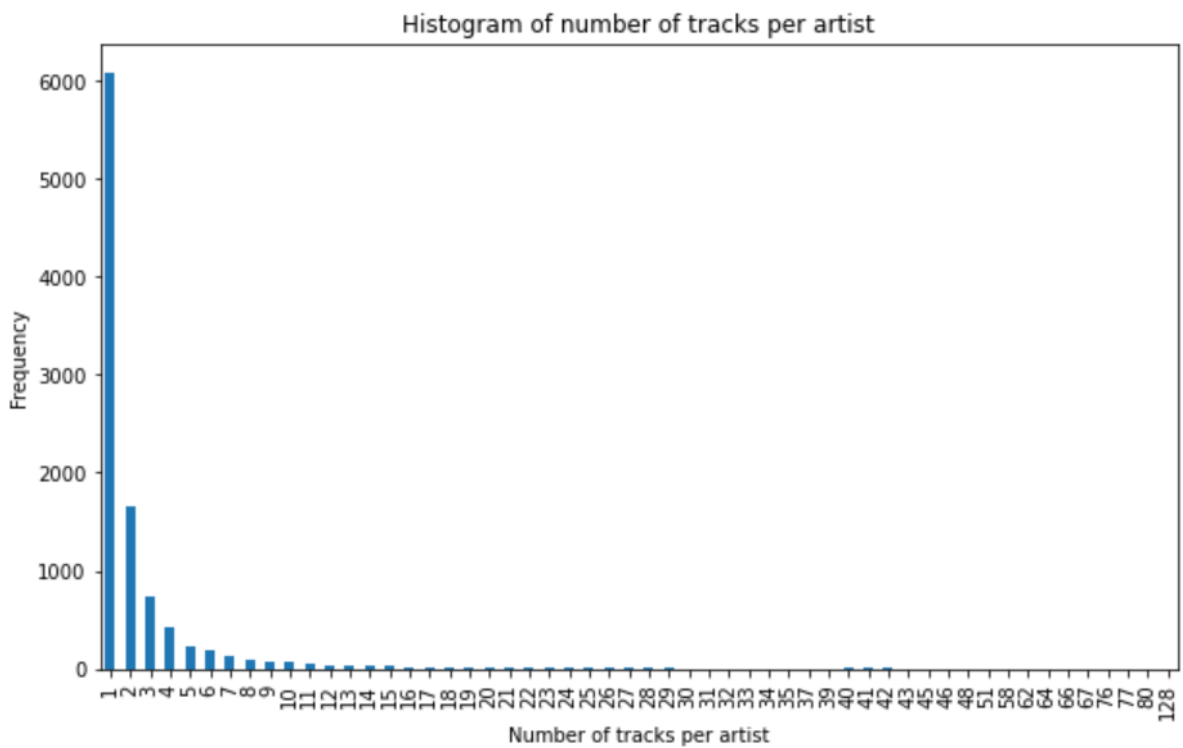

## Methods

In this section, we want to briefly discuss the ideas of different methods and discuss, why each of these is useful in this context and what may drawbacks of these methods.

### Ordinary Least Squares Regression (OLS)

OLS is widely employed in various fields, from economics to social sciences, for analyzing relationships between variables and making predictions based on historical data. Its strength lies in its simplicity, ease of implementation, and interpretability. However, OLS has some weaknesses, such as its sensitivity to outliers, assumptions of linearity, independence, homoscedasticity, and normality of residuals. Violations of these assumptions can lead to biased parameter estimates and inaccurate predictions, necessitating caution and supplementary diagnostics when applying OLS regression. Moreover, it is not very good in describing multicollinear variables and models with many categorical variables can get, complex very fast.

## Multivariate Adaptive Regression Splines (MARS)

MARS, a nonparametric method for multiple regression, uses adaptively selected spline functions to model high variation and smoothness in predictor space (Kooperberg, 2001). It is a flexible regression modelling method for high-dimensional data, automatically determining the number of basic functions and their parameters (Friedman, 1991). In other words, MARS offers more flexible models than OLS regressions as it allows the variables to be better explained using hinge functions. Instead of simply drawing a straight line through the data, the MARS model allows the lines to be bent at one or more points or to ignore sections of a variable that do not contribute to improving the predictive quality of the model. Researchers have found out that Linear models for song popularity prediction perform poorly when compared to models that can capture non-linearity (Gulmatico et al. 2022).

## Random forest regression

Random forests, also known as random decision forests, are a type of ensemble learning technique utilized for classification, regression, and similar tasks. This method involves creating numerous decision trees during the training phase. In classification scenarios, the random forest's output is determined by the class most frequently chosen across the trees. In regression tasks, the collective prediction from the individual trees is typically expressed as the mean or average (Tin Kam Ho, 1995).

## XGBoost

The Gradient Tree Boosting Algorithm is an ensemble learning method designed for both classification and regression tasks, much like the random forest algorithm. Ensemble learning involves combining multiple machine learning algorithms to improve model performance. The concept of "gradient boosting" revolves around the idea of improving a weak model by combining it with several others to create a stronger collective model (Chen & Guestrin, 2016).

In gradient boosting, the objective function for subsequent models is determined to minimize errors, with each case's targeted outcome based on the gradient of the error concerning the prediction. Gradient boosting decision trees (GBDT) iteratively train a set of shallow decision trees, with each iteration utilizing the error residuals from the previous model to refine the next one. The final prediction is a weighted sum of all tree predictions. While random forest employs bagging to mitigate variance and overfitting, GBDT utilizes boosting to reduce bias and underfitting (Nvidia, 2024).

## Results

### OLS

| | coef | std err | t | P>|t| | [0.025 | 0.975] |
|---|---|---|---|---|---|---|
| *Intercept* | 55.2674 | 1.05 | 52.633 | 0 | 53.209 | 57.326 |
| *C(playlist_genre, Treatment(reference="rock"))[T.edm]* | -50.0955 | 6.575 | -7.619 | 0 | -62.984 | -37.207 |
| *C(playlist_genre, Treatment(reference="rock"))[T.latin]* | -19.4239 | 3.339 | -5.818 | 0 | -25.968 | -12.88 |
| *C(playlist_genre, Treatment(reference="rock"))[T.pop]* | -4.2928 | 2.825 | -1.519 | 0.129 | -9.831 | 1.245 |
| *C(playlist_genre, Treatment(reference="rock"))[T.r&b]* | -22.5725 | 1.949 | -11.583 | 0 | -26.392 | -18.753 |
| *C(playlist_genre, Treatment(reference="rock"))[T.rap]* | -29.44 | 2.446 | -12.035 | 0 | -34.235 | -24.645 |
| *C(mode)[T.1]* | 0.0719 | 0.29 | 0.248 | 0.804 | -0.497 | 0.641 |
| *danceability* | 0.9319 | 0.177 | 5.278 | 0 | 0.586 | 1.278 |
| *energy* | -3.2299 | 0.243 | -13.314 | 0 | -3.705 | -2.754 |
| *loudness* | 3.4024 | 0.219 | 15.512 | 0 | 2.972 | 3.832 |
| *speechiness* | -0.1204 | 0.162 | -0.744 | 0.457 | -0.437 | 0.197 |
| *acousticness* | 0.3441 | 0.173 | 1.992 | 0.046 | 0.005 | 0.683 |
| *instrumentalness* | -1.4724 | 0.157 | -9.379 | 0 | -1.78 | -1.165 |
| *liveness* | -0.1642 | 0.144 | -1.139 | 0.255 | -0.447 | 0.118 |
| *valence* | 0.0739 | 0.17 | 0.434 | 0.664 | -0.26 | 0.407 |
| *tempo* | 0.5581 | 0.146 | 3.812 | 0 | 0.271 | 0.845 |
| *duration_ms* | -1.1643 | 0.156 | -7.451 | 0 | -1.471 | -0.858 |
| *years_since_1957* | -0.2241 | 0.023 | -9.916 | 0 | -0.268 | -0.18 |
| *years_since_1957:C(playlist_genre, Treatment(reference="rock"))[T.edm]* | 0.7539 | 0.111 | 6.817 | 0 | 0.537 | 0.971 |
| *years_since_1957:C(playlist_genre, Treatment(reference="rock"))[T.latin]* | 0.3691 | 0.059 | 6.207 | 0 | 0.253 | 0.486 |
| *years_since_1957:C(playlist_genre, Treatment(reference="rock"))[T.pop]* | 0.1754 | 0.051 | 3.432 | 0.001 | 0.075 | 0.276 |
| *years_since_1957:C(playlist_genre, Treatment(reference="rock"))[T.r&b]* | 0.3567 | 0.039 | 9.177 | 0 | 0.281 | 0.433 |
| years_since_1957:C(playlist_genre, Treatment(reference="rock"))[T.rap] | 0.5585 | 0.045 | 12.397 | 0 | 0.47 | 0.647 |
| key | 0.126 | 0.143 | 0.881 | 0.378 | -0.154 | 0.406 |

*Table 1: Coefficients of OLS-Regression Analysis with Genre and time influence*

The OLS regression analysis reveals some interesting correlations. By standardising all numerical variables, the coefficients can be compared with each other and the strength of the influence of these variables can be interpreted easily. The variable *years_since_1957* is the only one that was not standardised and ranges from 0, which corresponds to the year 1957, to 63, which corresponds to the year 2020. The model was trained using 20588 observations and has an $R^2$ of 8.7% which is rather small to moderate.

We can observe that the genre has the greatest effect on song popularity. While the genre Rock has its highest popularity values in 1957, the average popularity of the genre decreases by 0.2 popularity points every year. From an average of 55 popularity points in 1963, this value drops by an average of 14 points to 41 points in 2020. The edm genre, which was only a niche genre in 1963, has seen the strongest popularity gain in the period up to 2020. While its popularity was only 4 in 1963, its popularity score has risen to 52 over time. The genres, which have gained in popularity over the years are here presented in descending order: edm, latin, r&b, pop. It is important to emphasise that pop has already enjoyed a high level of popularity since 1963, only just below that of the rock genre, and has since then only gained in popularity.

All other variables (mode, danceability, energy, loudness, speechiness, acousticness, instrumetalness, liveness, valence, tempo and duration) tend to play a subordinate role in popularity in comparison. These variables do not have such a strong as the genre has. Nevertheless, it can be observed that, at least in the genre edm, energy has the strongest negative effect (-3.2) and loudness the strongest positive effect (3.4). So, if a song has a higher energy, the song becomes less popular on average. One standard deviation more energy reduces the popularity score by 3.2 points. It should be noted, of course, that a standard deviation difference is relative. If the songs have a very high variance in energy, then one standard deviation is a lot (e.g. one standard deviation can have a difference in energy level of 0.3), if the variance is very small, then one standard deviation difference is very little in terms of absolute difference in the energy level (e.g. one standard deviation can have a difference in energy level of 0.05).

If we want to cast a cross-genre and cross-time influence of the variables mentioned in the last paragraph, we need a genre-independent model. These coefficients are shown in Table 2. The model also has 20588 observations but an $R^2$ of only 4.6%. Without the genre and time components, this model explains significantly less of the variance. Therefore, the previous model is preferable. Nevertheless, the analysis of this model is interesting as it allows us to take a closer look at the other variables in a cross-genre context. This model shows very similar patterns to the model whose coefficients are shown in Table 1. As genres are represented in the data set with a similar frequency, the results are representative for all genres in the dataset. The model shows that the following variables have a positive influence: loudness (3.2), tempo (0.64), valence (0.64), acousticness (0.51). Negative variables are: energy (-3.61), instrumentalness (-2.06), duration (-1.75), liveness (-0.39) and speechiness (-0.35). The other variables are not significant at the 5% significance level.

| | coef | std err | t | P>|t| | [0.025 | 0.975] |
|---|---|---|---|---|---|---|
| Intercept | 43.1112 | 0.22 | 195.833 | 0 | 42.68 | 43.543 |
| C(mode)[T.1] | 0.5731 | 0.295 | 1.944 | 0.052 | -0.005 | 1.151 |
| danceability | 0.2947 | 0.164 | 1.792 | 0.073 | -0.028 | 0.617 |
| energy | -3.6125 | 0.236 | -15.279 | 0 | -4.076 | -3.149 |
| loudness | 3.2216 | 0.211 | 15.278 | 0 | 2.808 | 3.635 |
| speechiness | -0.3496 | 0.148 | -2.356 | 0.018 | -0.64 | -0.059 |
| acousticness | 0.5084 | 0.174 | 2.917 | 0.004 | 0.167 | 0.85 |
| instrumentalness | -2.057 | 0.153 | -13.473 | 0 | -2.356 | -1.758 |
| liveness | -0.3807 | 0.147 | -2.592 | 0.01 | -0.669 | -0.093 |
| valence | 0.6394 | 0.161 | 3.965 | 0 | 0.323 | 0.956 |
| tempo | 0.6422 | 0.148 | 4.337 | 0 | 0.352 | 0.933 |
| duration_ms | -1.7498 | 0.148 | -11.852 | 0 | -2.039 | -1.46 |
| key | 0.1187 | 0.146 | 0.814 | 0.416 | -0.167 | 0.405 |

*Table 2 Coefficients of OLS-Regression Analysis without Genre and time influence*

## MARS

The summary of the MARS model can be found in Table 3. To enable a better comparison with the OLS model, it was also carried out without interaction effects.

| Basis Function | Pruned | Coefficient |
|---|---|---|
| (Intercept) | No | 28.808 |
| h(instrumentalness+0.380409) | No | -1.24926 |
| h(-0.380409-instrumentalness) | No | 910.514 |
| h(duration_ms+2.39678) | No | -1.69278 |
| h(-2.39678-duration_ms) | No | -33.4942 |
| h(energy-1.64918) | Yes | None |
| h(1.64918-energy) | No | 3.91307 |
| h(loudness+2.25395) | No | 3.57536 |
| h(-2.25395-loudness) | Yes | None |
| h(valence+2.03795) | No | 0.725436 |
| h(-2.03795-valence) | No | 262.584 |
| h(tempo+2.22622) | No | 0.557099 |
| h(-2.22622-tempo) | No | -73.0703 |
| h(loudness-1.48399) | No | -17.5395 |
| h(1.48399-loudness) | Yes | None |

In the model summary it can be observed that 2 hinge functions have not been included in the model. These are subspaces from the variables energy and loudness. Loudness has an additional breakpoint compared to the other variables. From this it can be suspected that the relationship between loudness and popularity is more complex, than these of the other variables. The MARS algorithm finds a better fit of the model, when this variable is further splitted into subsections, including more hinge functions for it. This indicates that the term loudness has a non-linear relationship with the popularity of the title, whereby both high and low loudness values can influence the popularity of the title differently.

## Random Forest

To interpret the meaning of the integrated variables, we can use the criterion of feature meaning. The Python implementation of Random Forests offers the possibility to compare the features based on the impurity of the variables of each tree. The importance of a trait is calculated as the (normalised) total reduction of the criterion by that trait. It is also known as the Gini value. (SciKit, 2024)

Impurity-based importance has a bias towards features with high cardinality, as higher cardinality tends towards finer-grained leaves and nodes. Therefore, leaves of higher cardinality will automatically have higher purity. For this reason, the diagram should be considered in blocks of different cardinality. There is the first block with cardinality of 2: genres and modes (cardinality of 2, because values can either be 0 or 1). Each of the importance score of the variables in the green cluster can be compared directly with each other. The values of the variables of the blue cluster are all in between 0 and 1 and are therefore also comparable. The grey-coloured variables have very different. Therefore the importance of each of these variables cannot be directly compared.

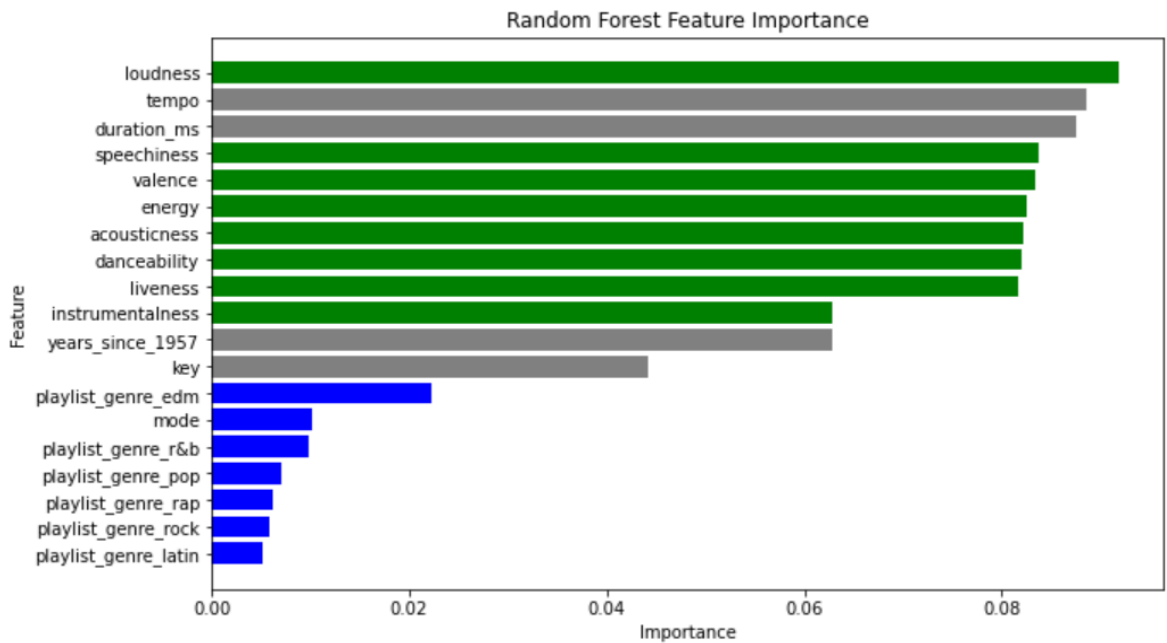

*Figure 1 Feature importance of Random Forest*

Striking is that the feature importance of the variables relating to the song characteristics (coloured green) have a similar importance. However, with the exception that loudness has the greatest influence and instrumentalness by far the smallest. In the OLS model, loudness also has a high importance and instrumentalness is insignificant. Among the genres (coloured in blue), the genre edm has the highest importance. This means that among the genres, in edm it is the easiest to determine popularity. For the genre latin, prediction of song popularity is the most difficult.

## XGBoost

The interpretation of the attribute importance of the XGBoost model is analogous to the Random Forest model. Here, the importance for an individual decision tree is calculated based on the amount by which each attribute split improves the prediction of song popularity. The difference is that each node is weighted according to the number of observations for which it is responsible. Therefore, a high cardinality does not distort the results. The performance measure can be the purity (Gini index) used to select the split points or another more specific error function. The feature values are then averaged across all decision trees within the model. (Brownlee, 2020)

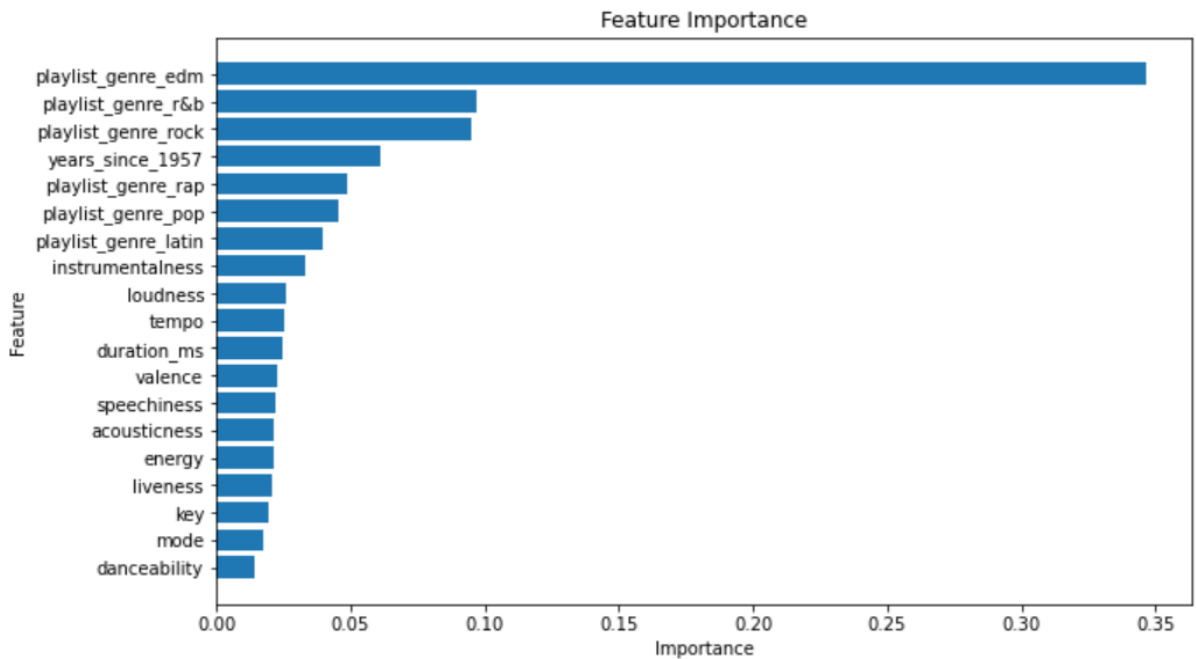

*Figure 2 Feature importance of XGBoost*

Figure 2 shows, in line with the OLS model, that genres, especially edm, are the most important in predicting the popularity of a song and latin is the least important. Interestingly, in contrast to the random forest model, the instrumentality of a song is most important. Loudness is also considered important here. A similar pattern can be observed in all other models for the least important characteristic of danceability.

## Comparison

For each model, the song popularity was predicted for the test data set. For all models, the mean absolute error (mae) was calculated and then compared with the mae of the mean song popularity. To get the mean song popularity, simply the average song popularity score of the training data set was calculated. The mae of the prediction of the song popularity using the average song popularity of the training data, which is 43.5, the mae is 17.55 points. This will serve as a benchmark and the improvements of all other models are compared to this mae.

The OLS full model has a mae of 16.64 and is therefore 5.2% better at predicting popularity. The MARS model shows an improvement of about 3 %, but does not include all the features that are present in the full OLS model. For a better comparison, a version of the OLS model that does not include the genre of the songs were included. The OLS model without genre performs slightly worse than the MARS model. The ability to increase the customisation of the model using the hinge functions improves it slightly, but the full OLS model (with genre) is clearly better than the MARS model. Including the genres as explanatory variables improves the model far more than adding hinge functions to the model.

The XGBoost predictions have better predictive power than the full ols model. With its 16.3 mae, it makes 6.2 % fewer errors on average. Unsurprisingly, the model performs better at predicting popularity for the training data, but the fact that it also performs better on the test set shows that although the XGBoost model tends to overfit more than other models in general, it is better in this scenario.

The Random Forest model was the best model for predicting the popularity of songs. It improves by 7.1% compared to the average popularity.

|  | mae_of_predicted_value | Improvement_in_mean |
|---|---|---|
| mean_track_polularity | 17.55 | 0.00 % |
| ols_full_model | 16.64 | 5.2 % |
| ols_wo_genre_model | 17.05 | 2.86 % |
| mars_wo_genre_model | 17.02 | 3.03 % |
| rf_predictions | 16.31 | 7.08 % |
| xgboost_predictions | 16.47 | 6.16 % |

## Discussion

Our study yielded valuable insights by successfully differentiating between various variables and identifying those that appear to exert more significant influence.

Through our analysis, we were able to discern patterns and relationships within the data, shedding light on the relative importance of different features in predicting song popularity. This process allowed us to identify key variables that play a prominent role in shaping audience preferences and determining the success of songs.

However, it is important to acknowledge a potential limitation in our study regarding the control for individual artists. The identity of the artist performing a song is undeniably relevant in determining its popularity, as artists often have distinct styles and fan bases. However, within our dataset, each artist is represented by a relatively small number of songs, averaging only 2.5 songs per artist. In such a scenario, employing a fixed effects model to account for individual artist effects would not be feasible due to insufficient data points per artist.

As a result, the influence of individual artists remains somewhat of a black box in our analysis. Failure to adequately account for artist-specific variables may lead to unexplained variability in our predictive models. Recognizing and incorporating artist-specific factors could potentially enhance the explanatory power ($R^2$) of our models by capturing additional sources of variation in song popularity.

Future research endeavours could explore strategies to address this limitation, such as incorporating external datasets containing more comprehensive information on individual artists, their discographies, and their respective fan bases. By elucidating the role of individual artists in shaping song popularity, we can further refine our predictive models and deepen our understanding of the complex dynamics underlying musical success.

## Conclusion & Implications

Some variables have clearly a high impact in predicting song popularity. The genre plays a big role, especially edm. It has moreover gained a lot of popularity from 1963 until 2020. This also shows that the music taste is changing over time. Rock, used to be the most popular genre, but it has decreased a quite substantial popularity. Music related features do also play a role in predicting song popularity, but as most model suggested, not that much than the genre. As discussed in Section 6, the music related features are of course intertwined with the genres, so it is not trivial to separate the effects between the features and the genres. However, with relatively high confidence we can state that loud songs, relatively short songs tend to rank higher in popularity scores.

Moreover, it's imperative to acknowledge the significant impact of post-release variables, such as marketing strategies and social media reception, in shaping song popularity. These factors, which are not captured in our predictive models, can significantly influence a song's trajectory in the public sphere. It's important to recognize that popularity does not necessarily equate to quality, as the posteriori variables can heavily influence perception and reception.

The fascination and relentless pursuit of unraveling the elusive magic formula behind song popularity will undoubtedly continue to captivate researchers in the future. While the quest to decode this phenomenon remains ongoing, the role of machine learning algorithms in this pursuit cannot be overstated. These algorithms serve as powerful tools for analysing vast datasets and uncovering patterns that may elude traditional analytical approaches. As technology advances and datasets grow larger and more diverse, we remain keen to spectate whether a definitive solution to predicting song popularity can be achieved, and the extent to which machine learning algorithms will play a pivotal role in this enigma.

# Acknowledgements

This study was conducted during a Machine-Learning Course, a data science course of the *Digital Science Center* Innsbruck, an elective interdisciplinary module for master students. We thank Dr. Marica Valente for the inspiring discussions and the DiSC for their superb Teaching.